\documentclass{article}

\usepackage{epsf,hyperref}
\usepackage{amssymb,ComplexSystems}
\usepackage{physics}
\usepackage{graphicx}

\newtheorem{conjecture}{Conjecture}

\begin{document}

\title{Quantum Cellular Automata, Black Hole Thermodynamics, and the Laws of Quantum Complexity}

\author{\authname{Ruhi Shah}\\[2pt] 
\authadd{University of Waterloo}\\
\authadd{200 University Ave W, Waterloo, ON N2L 3G1, Canada}\\
\authadd{\url{r57shah@uwaterloo.ca}}\\
\and
\authname{Jonathan Gorard}\\[2pt]
\authadd{King's College, University of Cambridge}\\
\authadd{CB2 1ST, Cambridge, England}\\
\authadd{\url{jg865@cam.ac.uk}}\\[8pt]
\authadd{Algorithms R\&D Group, Wolfram Research, Inc.}\\
\authadd{100 Trade Center Dr, Champaign, IL 61820-7237, USA}\\
\authadd{\url{jonathang@wolfram.com}}
}

\markboth{Quantum Cellular Automata} 
{The Laws of Quantum Complexity} 
\maketitle

\begin{abstract}
This paper introduces a new formalism for quantum cellular automata (QCAs), based on evolving tensor products of qubits using local unitary operators. It subsequently uses this formalism to analyze and validate several conjectures, stemming from a formal analogy between quantum computational complexity theory and classical thermodynamics, that have arisen recently in the context of black hole physics. In particular, the apparent resonance and thermalization effects present within such QCAs are investigated, and it is demonstrated that the expected exponential relationships between the quantum circuit complexity of the evolution operator, the classical entropy of the equilibrium QCA state, and the characteristic equilibration time of the QCA, all hold within this new model. Finally, a rigorous explanation for this empirical relationship is provided, as well as for the relationship with black hole thermodynamics, by drawing an explicit mathematical connection with the mean ergodic theorem, and the ergodicity of $k$-local quantum systems.
\end{abstract}

\section{Introduction}
\label{intro}

When attempting to make a physical theory computable, a very natural first step is to discretize the underlying spacetime\cite{farrelly}\cite{feynman}\cite{tong}. When one performs such a discretization on an arbitrary quantum mechanical system obeying special relativity (i.e. one in which there exists a hard upper-bound on the rate of information propagation), one obtains a \textit{quantum cellular automaton}, or QCA. Though originally proposed as an alternative to quantum Turing machines as a canonical model for universal quantum computation\cite{arrighi}\cite{watrous}, QCAs have also been studied as models for computable quantum field theories, amongst other things\cite{wilson}.

On the other hand, quantum computational complexity (as measured in terms of circuit complexity, i.e. the minimal number of quantum gates required to prepare a specified unitary operator) has recently been studied extensively in the context of black hole complementarity and firewalls\cite{brown}\cite{susskind}, as well as the more general ${ER = EPR}$ conjecture of quantum gravity\cite{stanford}\cite{susskind2}. Here, it is hypothesized that both quantum circuit complexity and classical entropy should obey identical growth conditions\cite{brown2}. More specifically, it has been conjectured that there should exist some kind of analog of the second law of thermodynamics for quantum circuit complexity\cite{susskind3}, as this would ensure that any naturally-formed black hole would have a monotonically increasing quantum complexity, and therefore would maintain a transparent, firewall-free, horizon for an exponentially long time\cite{almheiri}.

The present paper introduces a new model for QCAs, based on evolving tensor products of qubits using local unitary operators, and uses it to analyze and validate a variety of these conjectures in quantum complexity theory, at least in toy cases. We show, in particular, that such QCAs exhibit definite ``thermalization'' effects; namely, we show that all such QCAs will eventually reach a stable equilibrium state, that the time taken to reach this state is independent of the initial conditions, and that this ``thermalization'' time is negatively exponentially related to the quantum circuit complexity, as one would expect by analogy to classical statistical mechanics. We also investigate the resonance effects inherent to this class of QCAs, and provide some explicit mathematical connections with ergodic theory, and with $k$-local systems from black hole physics.

\section{Mathematical Background}

\subsection{Quantum Cellular Automata}

In what follows, we will be employing the general mathematical formalism for QCAs developed and espoused by Arrighi et al.\cite{arrighi2}\cite{arrighi3}\cite{arrighi4}. Intuitively, a QCA is any finite collection of $n$ quantum systems, each of the same dimensionality, $d$; that is, each of the ${\mathbb{Z}^n}$ cells in the QCA is a qudit in the Hilbert space ${\mathcal{H}_d}$. One might therefore assume that the overall space of QCA states has the form:

\begin{equation}\nonumber
\mathcal{H} = \bigotimes_{\mathbb{Z}} \mathcal{H}_d,
\end{equation}
but this is not a Hilbert space in general (i.e. it may be a non-Hilbert ${C^*}$-algebra)\cite{schumacher}. Thus, one must be somewhat careful regarding precisely how the set of configurations is defined.

\begin{definition}
The \textit{alphabet} of a QCA, denoted ${\Sigma}$, is any finite set with a distinguished element, denoted $0$, namely the \textit{empty state}.
\end{definition}

\begin{definition}

A \textit{configuration} over ${\Sigma}$, denoted $c$, is any function:

\begin{equation}\nonumber
c : \mathbb{Z}^n \to \Sigma,
\end{equation}
such that the set:

\begin{equation}\nonumber
A = \left\lbrace \left( i_1, \dots, i_n \right) \in \mathbb{Z}^n \textit{ such that } c_{i_1 \dots i_n} \neq 0 \right\rbrace,
\end{equation}
is finite.
\end{definition}

The set of all such configurations, denoted ${\mathcal{C}}$, is countable, thus allowing us to define a Hilbert space of (superpositions of) configurations.

\begin{definition}
The \textit{state space} of configurations of the QCA, denoted ${\mathcal{H}}$, is the Hilbert space with canonical orthonormal basis:

\begin{equation}\nonumber
\left\lbrace \ket{c} \right\rbrace_{c \in \mathcal{C}}.
\end{equation}
\end{definition}

Finally, we must enforce two constraints on the global evolution of the QCA: translation-invariance, meaning that the evolution operator acts everywhere equally, and causality, meaning that there exists a definite upper-bound on the rate of propagation of information.

\begin{definition}
The \textit{translation operator} along the ${k^{th}}$ dimension, denoted ${\tau_k}$, is the linear operator over ${\mathcal{H}}$ mapping ${\ket{c}}$ to ${\ket{c^{\prime}}}$, where the ${\ket{c^{\prime}}}$ denotes the state such that:

\begin{equation}\nonumber
\forall \left( i_1, \dots, i_n \right), \qquad c^{\prime}_{i_1 \dots i_k \dots i_n} = c_{i_1 \dots i_{k + 1} \dots i_n}.
\end{equation}
\end{definition}

\begin{definition}
A \textit{translation-invariant} operator, denoted $G$, is any linear operator over ${\mathcal{H}}$ such that:

\begin{equation}\nonumber
\forall k, \qquad G \tau_k = \tau_k G
\end{equation}
\end{definition}

To formalize the statement that the QCA satisfies the causality requirement, we must first define a neighborhood structure: we want to be able to say that the state of a cell at time ${t + 1}$ should depend only upon the states of its neighbours at time $t$. However, in order to be able to speak meaningfully about states of subsystems of the QCA, we must introduce a density matrix formalism.

\begin{definition}
A \textit{density matrix}, denoted ${\rho}$, represents the set of probability distributions over pure states:

\begin{equation}\nonumber
\left\lbrace p_i, \ket{\psi_i} \right\rbrace,
\end{equation}
as a convex sum of projectors:

\begin{equation}\nonumber
\rho = \sum_i p_i \ket{\psi_i} \bra{\psi_i}.
\end{equation}
\end{definition}

Hence, if the pure states ${\ket{\psi}}$ evolve according to the rule:

\begin{equation}\nonumber
\ket{\psi^{\prime}} = G \ket{\psi},
\end{equation}
then the density matrices ${\rho}$ will evolve according to :

\begin{equation}\nonumber
\rho^{\prime} = G \rho G^{\dagger}
\end{equation}
For the purposes of the present paper, we will consider only idempotent density matrices, that is, cases in which the density matrix formalism reduces to the pure state vector formalism. Thus, the state of cell $x$ at time ${t + 1}$:

\begin{equation}\nonumber
x = \left( i_1, \dots, i_n \right),
\end{equation}
can be obtained by tracing out all other cells:

\begin{equation}\nonumber
\rho_{x}^{\prime} = \mathrm{Tr}_{\bar{x}} \left( \rho^{\prime} \right),
\end{equation}
where we have introduced the \textit{partial trace} linear operator, denoted ${\mathrm{Tr}_{\bar{S}} \left( \cdot \right)}$, defined by:

\begin{equation}\nonumber
\mathrm{Tr}_{\bar{S}} \left( \ket{c} \bra{d} \right) = \left( \delta_{c_{\bar{S}}, d_{\bar{S}}} \right) \ket{c_S} \bra{d_S}.
\end{equation}
In much the same way, the states of the neighbours of $x$ at time $t$ can be obtained by tracing out the remaining cells:

\begin{equation}\nonumber
\rho_{x + \mathcal{N}} = \mathrm{Tr}_{\bar{x + \mathcal{N}}} \left( \rho \right).
\end{equation}

\begin{definition}
A \textit{causal} operator, denoted $G$, with a neighborhod ${\mathcal{N} \subset \mathbb{Z}_n}$, is any linear operator over ${\mathcal{H}}$ such that:

\begin{equation}\nonumber
\exists f, \textit{ such that } \forall \rho \textit{ over } \mathcal{H}, \qquad \rho_{x}^{\prime} = f \left( \rho_{x + \mathcal{N}} \right),
\end{equation}
where, as usual:

\begin{equation}\nonumber
\rho^{\prime} = G \rho G^{\prime}.
\end{equation}
\end{definition}

\begin{definition}
A \textit{quantum cellular automaton} is any linear operator over ${\mathcal{H}}$ that is translation-invariant, causal, and unitary.
\end{definition}

\subsection{Quantum Complexity Theory}

So-called ``$k$-local systems'' provide a very natural and generic formalism for studying the dynamics of black holes. A $k$-local Hamiltonian is constructed from a sum of Hermitian operators, each containing at most $k$ qubits, such that that no operator has a weight higher than $k$. The ``weight'' of an operator, in this context, refers to the number of single-qubit factors which appear.

For instance, an ordinary QCA with a nearest-neighbor structure is an instance of a 2-local system, although it is important to note that $k$-locality does not imply spatial locality. An exactly $k$-local Hamiltonian, that is, one in which each operator acts on exactly $k$ qubits, has the general form:

\begin{equation}\nonumber
H = \sum_{i_1 < i_2 < \dots < i_k} \sum_{a_1 = \left\lbrace x, y, z \right\rbrace} \dots \sum_{a_k = \left\lbrace x, y, z \right\rbrace} J_{i_1, i_2, \dots, i_k}^{a_1, a_2, \dots, a_k} \sigma_{i_1}^{a_1} \sigma_{i_2}^{a_2} \dots \sigma_{i_k}^{a_k},
\end{equation}
or, in a more schematic form:

\begin{equation}\nonumber
H = \sum_I J_I \sigma_I,
\end{equation}
where $I$ runs over the set of all ${\left( 4^K - 1 \right)}$ generalized Pauli operators (i.e. the ${3K}$ Pauli operators ${\sigma_{i}^{a}}$, along with all possible products, without locality restrictions). Here, we assume that only $k$-local couplings between qubits are non-zero.

The work of Brown and Susskind concerns the question of how the quantum circuit complexity of the time-evolution operator:

\begin{equation}\nonumber
U(t) = e^{-i H t},
\end{equation}
for a general $k$-local system, evolves over time. The general form of the conjecture states that the complexity, denoted ${\mathcal{C} (t)}$, grows linearly:

\begin{equation}\nonumber
\mathcal{C} (t) = k t,
\end{equation}
for a period of time that is exponential in $k$. Thus, at time ${t \approx e^k}$, we reach the period of \textit{complexity equilibrium}, where complexity reaches its maximum value, denoted ${\mathcal{C}_{max}}$, and flattens out, fluctuating around the maximum:

\begin{equation}\nonumber
\mathcal{C}_{max} \approx e^k.
\end{equation}
Over much longer timescales, on the order of ${e^{e^k}}$, the complexity is assumed to return quasiperiodically to sub-exponential values, due to quantum recurrences.

These conjectures may be summarized succinctly, by making a formal analogy with classical statistical mechanics:

\begin{conjecture}
The quantum circuit complexity for a system of $k$ qubits behaves analogously to the entropy of a classical system with ${2^k}$ degrees of freedom.
\end{conjecture}

\section{The Quantum Tensor Automaton Model}

\subsection{Mathematical Formalism}

Within the new QCA formalism proposed in this paper, the overall state of the QCA at time $t$, denoted ${\ket{\psi^t}}$, is assumed to be given by a tensor product of the individual qubits at time $t$, denoted ${\ket{\phi_{j}^{t}}}$:

\begin{equation}\nonumber
\ket{\psi^t} = \ket{ \bigotimes_{j \in \mathbb{Z}_n} \phi_{j}^{t} }
\end{equation}
We also assume that the QCA begins with a pure initial state, ${\ket{\psi^0}}$. To evolve ${\ket{\psi^0}}$, we first extract the first qubit by tracing out all other cells, and we apply a unitary operator, denoted ${u^1}$, both to it and its immediate rightmost neighbor. This operation is performed for a sequence of operators ${u^1, \dots u^m}$, yielding the intermediate state ${\ket{\psi_{\prime}^{1}}}$, and then repeated for all qubits from 1 to $n$, yielding a family of intermediate states ${\ket{\psi_{1}^{\prime}}, \dots, \ket{\psi_{N}^{\prime}}}$. The intermediate states are then summed over to generate the next step, ${\ket{\psi^1}}$.

In other words, we can write the global evolution operator for the QCA, denoted $U$, as a ${2^n \times 2^n}$ unitary matrix of the form:

\begin{equation}\nonumber
U = \sum_{j \in \mathbb{Z}_n} \prod_{k \in \mathbb{Z}_m} u^{k}_{j} u^{k}_{j + 1}.
\end{equation}

We have implemented this formalism as a Wolfram Language function called \textit{QuantumTensorAutomaton}, which is currently available in the Wolfram Function Repository\cite{shah}. This function was used to produce all of the results presented within this paper.

\subsection{Some Initial Results}

\begin{figure}[ht]
\includegraphics[width=0.95\textwidth]{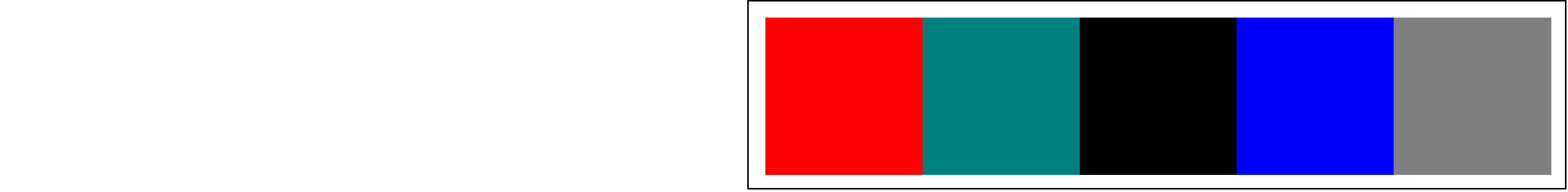}
\caption{Color scheme for visualization of complex numbers. The first square denotes $i$, the second denotes ${-i}$, the third denotes 0, the fourth denotes 1, and the fifth denotes -1. These colors are then blended to represent an arbitrary (normalized) complex number.}
\label{figure1}
\end{figure}

To visualize the evolution of QCAs (and, in particular, to visualize individual quantum amplitudes), we introduce the RGB visualization scheme for complex numbers shown in Figure \ref{figure1}.

\begin{figure}[ht]
\includegraphics[width=0.95\textwidth]{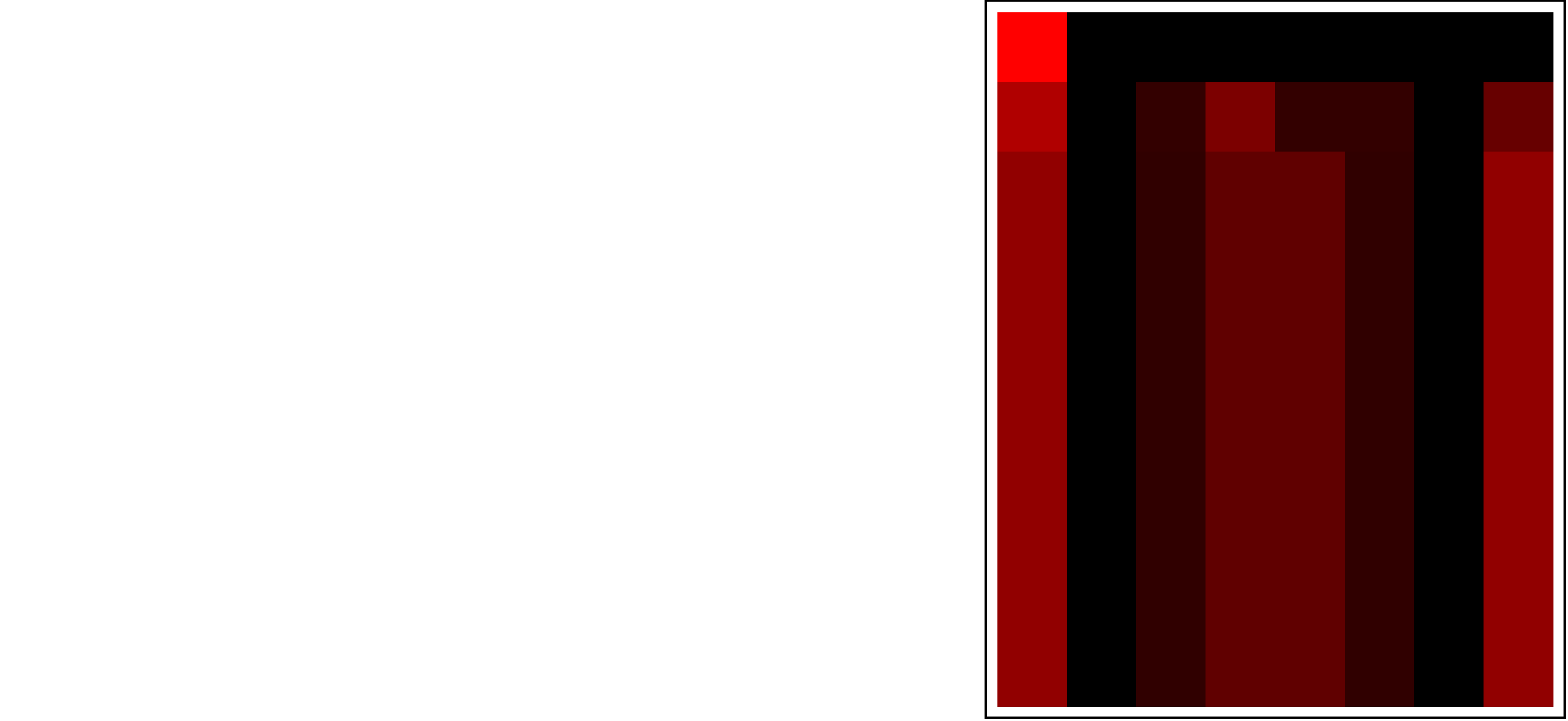}
\caption{The first 10 steps in the evolution of the CNOT QCA, given the canonical 3-qubit initial condition.}
\label{figure2}
\end{figure}

For instance, consider a QCA consisting of three qubits, with initial state vector ${(1, 0, 0, 0, 0, 0, 0, 0)}$, i.e:

\begin{equation}\nonumber
\ket{\psi^0} = 1 \ket{000} + 0 \ket{001} + 0 \ket{010} + 0 \ket{011} + 0 \ket{100} + 0 \ket{101} + 0 \ket{110} + 0 \ket{111},
\end{equation}
and evolved using the (arity 2) CNOT quantum logic gate:

\begin{equation}\nonumber
CNOT = \begin{bmatrix}
1 & 0 & 0 & 0\\
0 & 1 & 0 & 0\\
0 & 0 & 0 & 1\\
0 & 0 & 1 & 0
\end{bmatrix},
\end{equation}
which we can plot in a similar manner to a classical CA (i.e. with the state vector represented horizontally, and the evolution occurring down the page), as shown in Figure \ref{figure2}.

\begin{figure}[ht]
\includegraphics[width=0.95\textwidth]{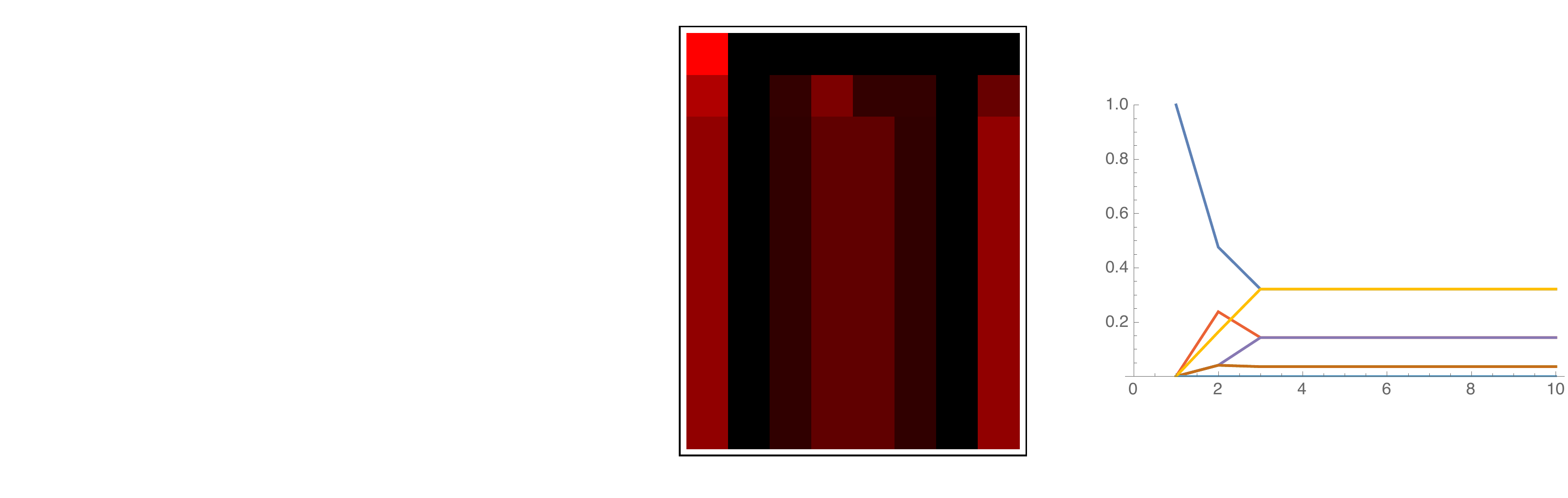}
\caption{On the left, the CNOT QCA evolved for 10 steps for the canonical 3-qubit initial condition. On the right, the associated probabilities for each basis state.}
\label{figure3}
\end{figure}

Rather than visualizing only the amplitudes, we can also plot how the associated probabilities (i.e. the norms of the amplitudes, squared) for each basis state change over time, in which case we immediately observe an equilibration effect, wherein the probabilities appear to stop changing after some finite time, as seen in Figure \ref{figure3}.

\begin{figure}[ht]
\includegraphics[width=0.95\textwidth]{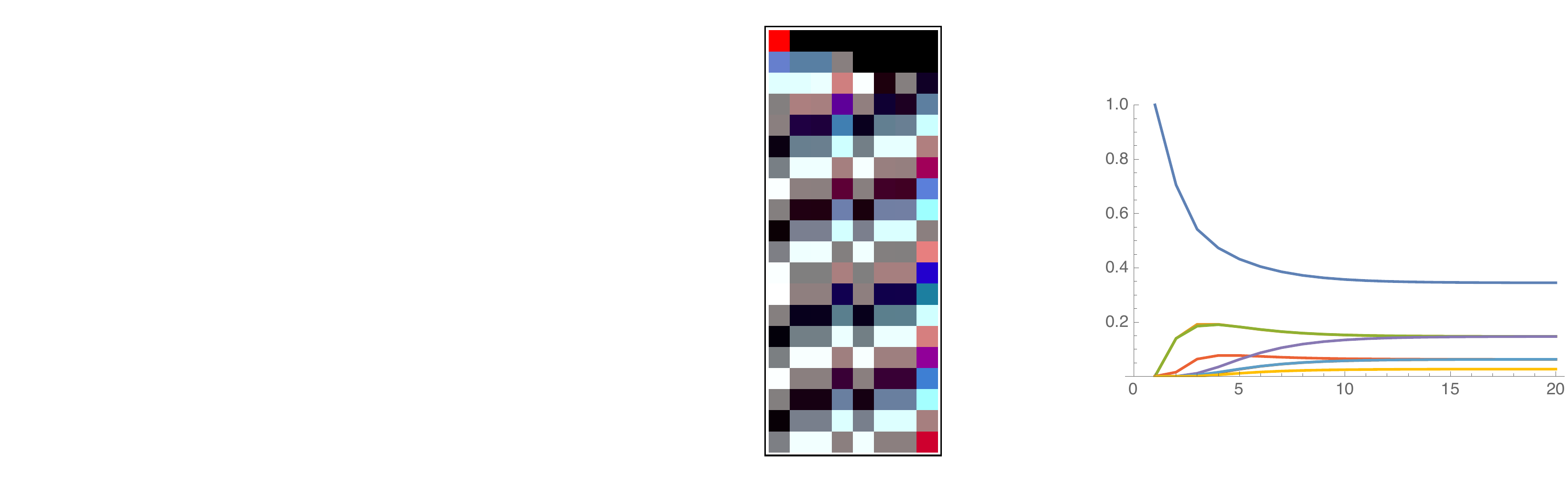}
\caption{Equilibration effects observed for a random unitary matrix operator of arity 2, over the course of 20 iterations.}
\label{figure4}
\end{figure}

In the case of the CNOT operator, this discovery is unsurprising, as the state vector itself reaches equilibrium after just a couple of iterations. However, the more interesting fact is that the same phenomenon can be observed for any arbitrary unitary matrix operator of arity 2, as shown in Figure \ref{figure4}.

\begin{figure}[ht]
\includegraphics[width=0.95\textwidth]{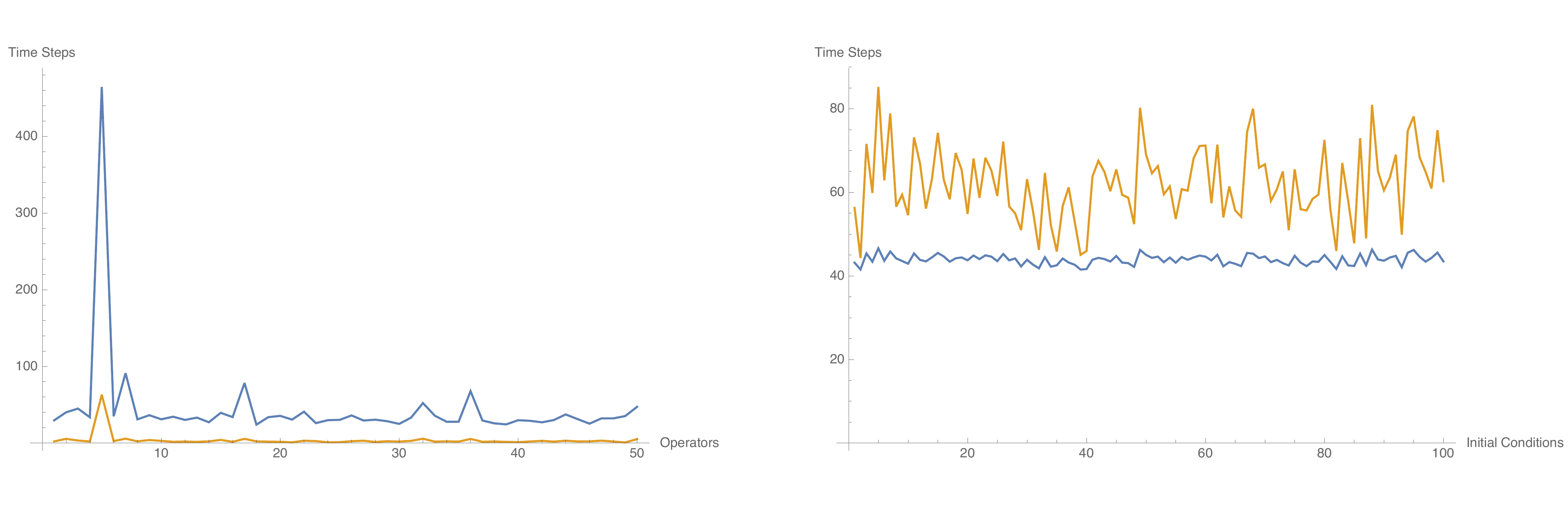}
\caption{The means (in blue) and standard deviations (in yellow) of the characteristic equilibration times. On the left, the average is taken over 100 random initial conditions, and plotted for 50 random unitary matrix operators. On the right, the average is taken over 50 random unitary matrix operators, and plotted for 100 random initial conditions.}
\label{figure5}
\end{figure}

Neither the equilibrium state, nor the characteristic time required to reach equilibrium, appear to be meaningfully affected by the choice of initial condition. To see this, we can compute the characteristic equilibration time for a given operator/initial condition pair, by determining the number of iterations required for the probabilities to stop changing by more than some fixed absolute value, ${\epsilon}$. As shown in Figure \ref{figure5}, the standard deviation of the characteristic equilibration time for a given unitary matrix operator, when averaged over 100 random initial conditions, is generally relatively low, whereas the standard deviation for a given initial condition, when averaged over 50 random unitary matrix operators, is generally relatively high. We can conclude that the equilibration times are indeed somewhat sensitive to the choice of operator, but are generally indifferent to the choice of initial condition.

\begin{figure}[ht]
\includegraphics[width=0.95\textwidth]{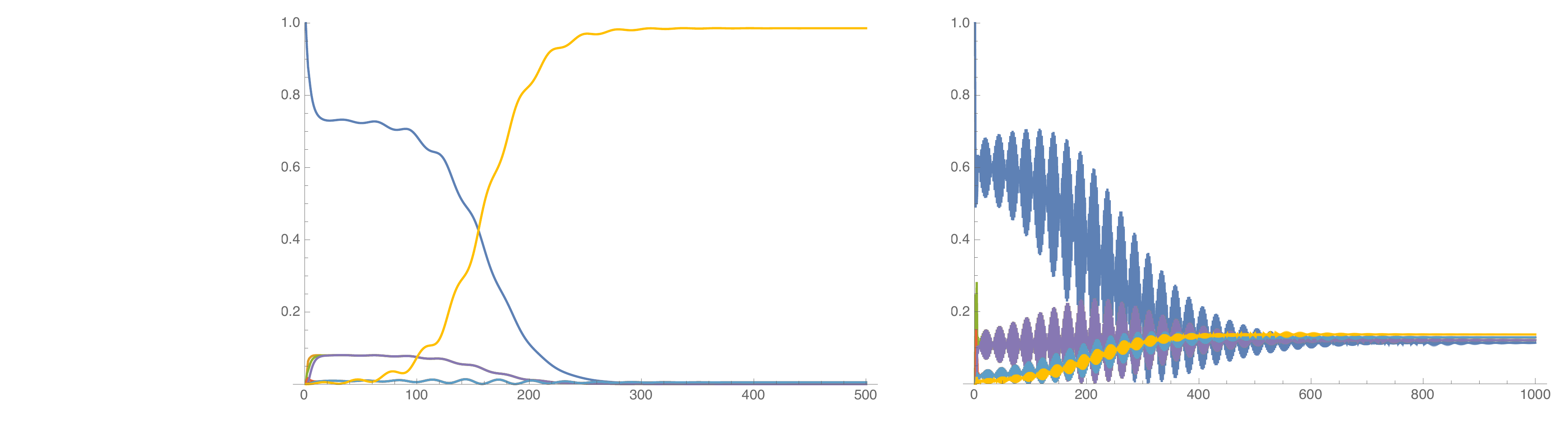}
\caption{Equilibration effects for two operators exhibiting resonance. On the left is the operator which created the spike in Figure \ref{figure5}, and on the right is an example of an operator exhibiting a much-enhanced version of the same phenomenon.}
\label{figure6}
\end{figure}

The spike observed on the left-hand-side of Figure \ref{figure5} is indicative of a unitary matrix operator with a much longer characteristic equilibration time than average. As shown in Figure \ref{figure6}, such operators may be thought of as exhibiting resonance effects.

\begin{figure}[ht]
\includegraphics[width=0.95\textwidth]{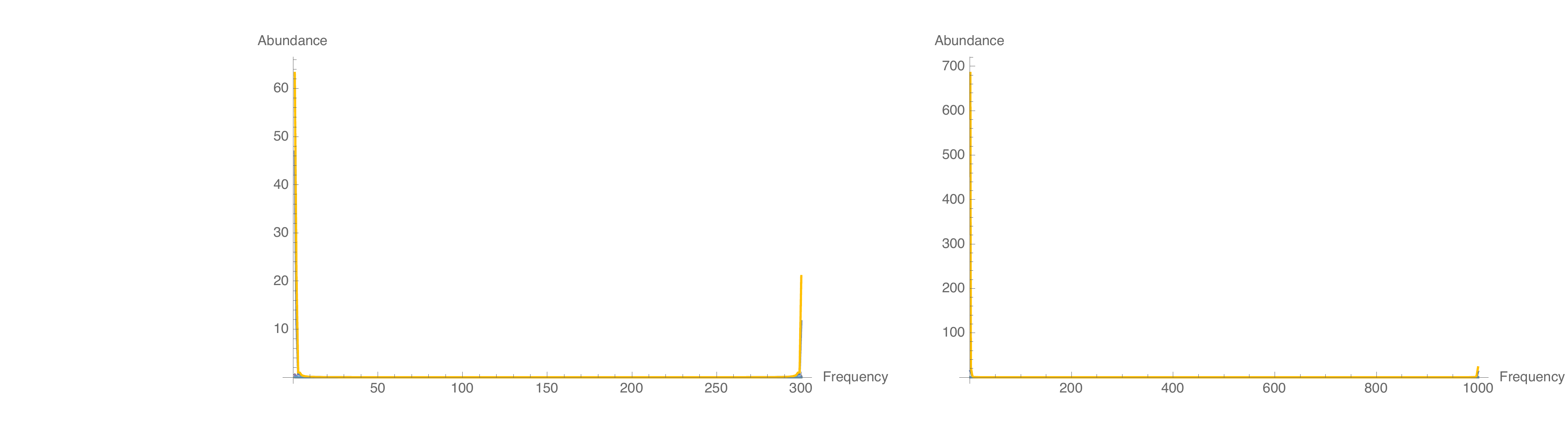}
\caption{Fourier plots showing the dominant frequencies for the resonant operator in Figure \ref{figure5}, first over 300 steps (left), and then over 1000 steps (right).}
\label{figure7}
\end{figure}

We can extract the resonant frequencies of such QCAs using techniques of Fourier analysis (i.e. by taking the discrete Fourier transform of the time evolution, and then plotting the norms of the Fourier coefficients, squared, to determine the dominant frequencies), as shown in Figure \ref{figure7}.

\begin{figure}[ht]
\includegraphics[width=0.95\textwidth]{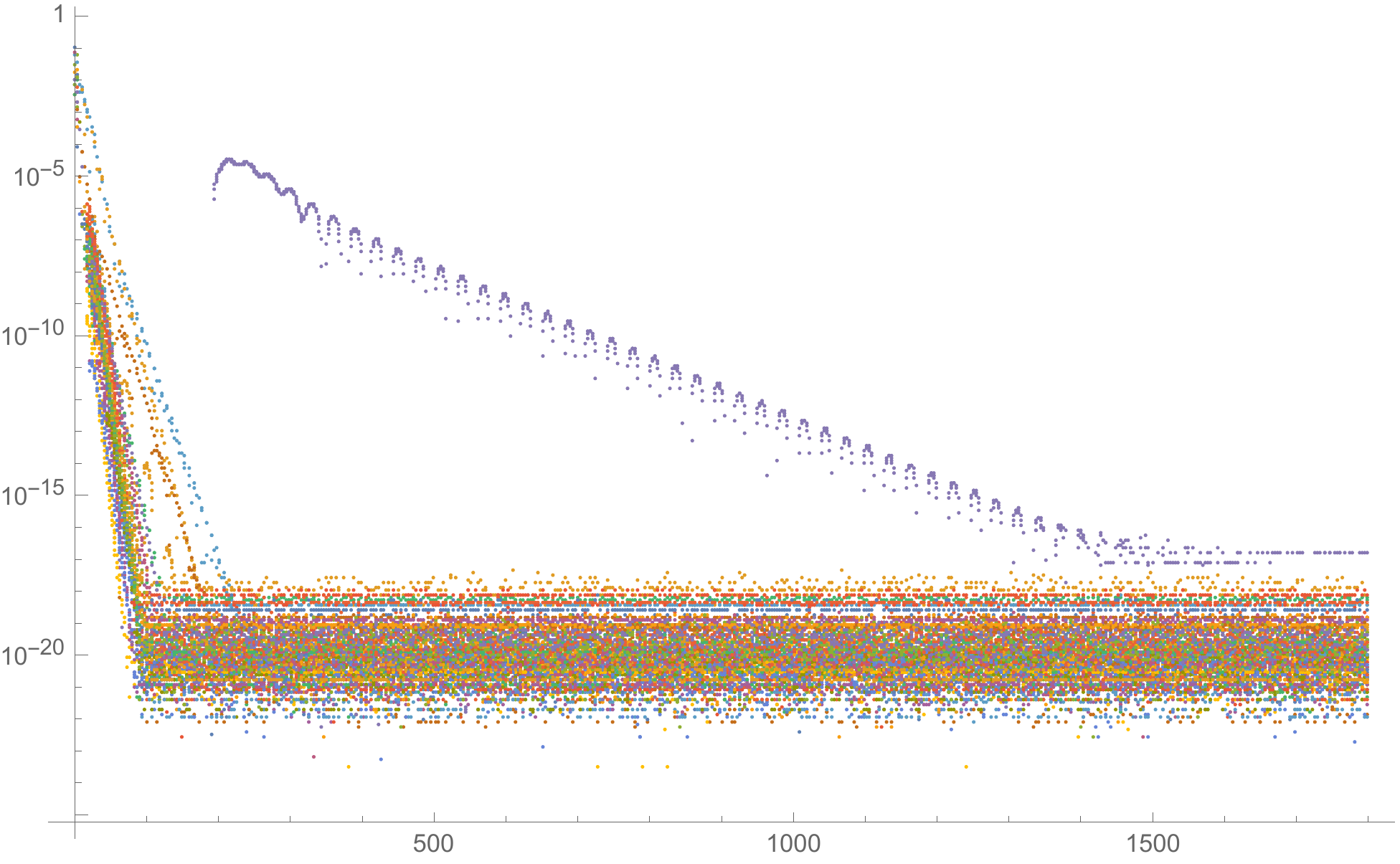}
\caption{A plot showing the convergence to equilibrium, with respect to the ${L^2}$-norm, of 50 random unitary matrix operators evolved over time (and averaged over 100 random initial conditions).}
\label{figure8}
\end{figure}

Since the resonant frequencies (i.e. those frequencies which are not close to zero) tend to die away as one increases the number of steps, we conclude that even the resonant operators will eventually reach an equilibrium state, only with characteristic timescales that happen to be much longer than average. This can be observed directly by performing a formal convergence analysis, with respect to the ${L^2}$-norm, of 50 random unitary matrix operators evolved over time (averaged over 100 random initial conditions), as shown in Figure \ref{figure8}.

\section{Quantum Complexity Theory}

\subsection{Approximating Circuit Complexity}

Even though the global evolution operator of a QCA is always unitary (and hence the evolution is always reversible), QCA evolution can, in practice, be irreducibly difficult to reverse. By analogy to the case of classical reversible CAs, one can think of this as being due to the QCA progressively ``encrypting'' the details of its initial condition as it evolves. Thus, one can recast the problem of determining the computational complexity of the time evolution operator into the problem of determining the quantum circuit complexity of the minimal ``reversal'' operator (i.e. the minimal unitary operator that correctly reconstructs the initial condition, given the evolved state of the QCA).

\begin{figure}[ht]
\includegraphics[width=0.95\textwidth]{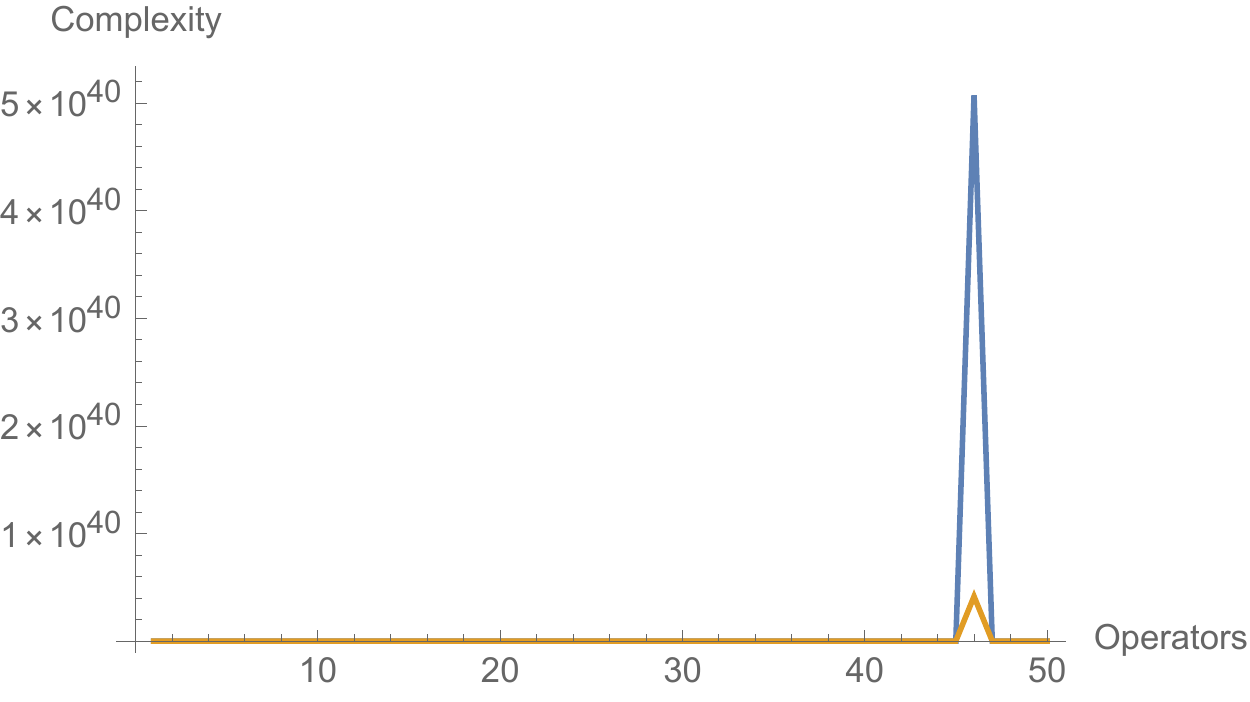}
\caption{The means (in blue) and standard deviations (in yellow) of the coarse-grained circuit complexity, averaged over 100 random initial conditions, and plotted for 50 random unitary matrix operators.}
\label{figure9}
\end{figure}

\begin{figure}[ht]
\includegraphics[width=0.95\textwidth]{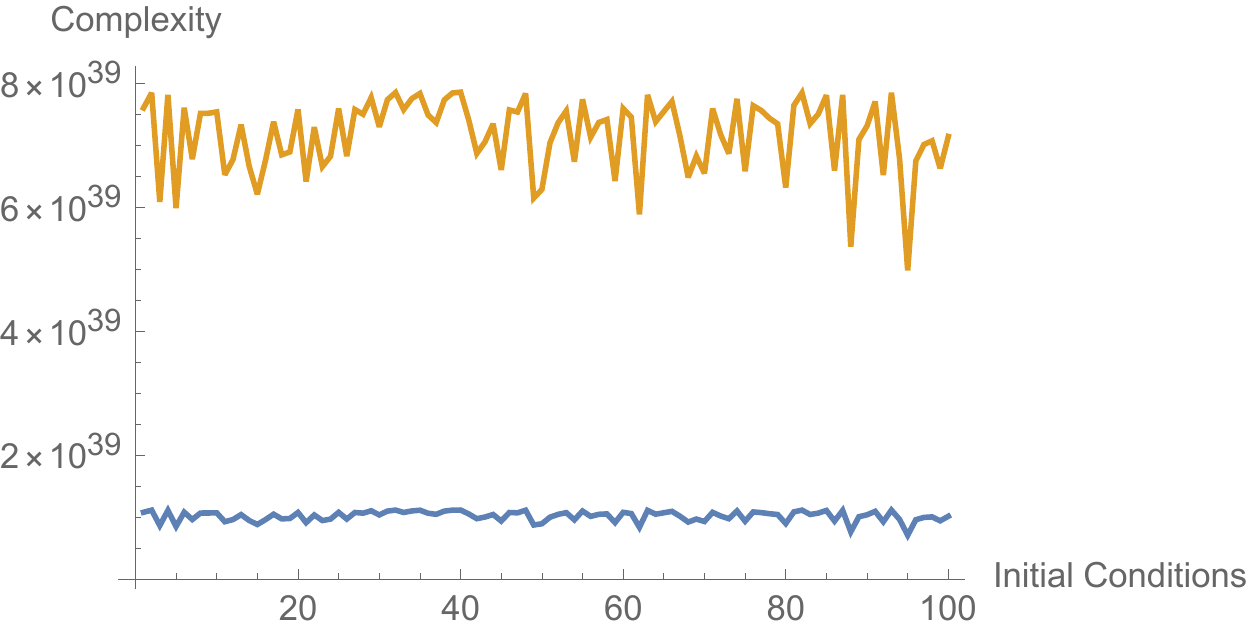}
\caption{The means (in blue) and standard deviations (in yellow) of the coarse-grained circuit complexity, averaged over 50 random unitary matrix operators, and plotted for 100 random initial conditions.}
\label{figure10}
\end{figure}

Thus, the reversal operator, $R$, at time $t$ can be computed easily, by finding the general matrix-valued solution of:

\begin{equation}\nonumber
R \ket{\psi^t} = \ket{\psi^0},
\end{equation}
and then finding a particular instance for which $R$ satisfies Hermiticity (in effect, we are treating the time-reversal process as being analogous to the measurement of some observable). Then, one can compute a coarse-grained approximation to the circuit complexity of $R$ by summing over the squared moduli of its eigenvalues. We see, in Figures \ref{figure9} and \ref{figure10}, that the statistical behavior of the coarse-grained circuit complexity is extremely similar to that of the characteristic equilibration times; the standard deviation of the complexity for a given unitary matrix operator, averaged over 100 random initial conditions, is generally relatively low, whereas the standard deviation for a given initial condition, when averaged over 50 random unitary matrix operators, is generally relatively high. Thus, just as with equilibration times, circuit complexity appears to be fairly sensitive to the choice of operator, but is more-or-less unaffected by the choice of initial condition.

Indeed, performing a nonlinear model fit between the natural logarithms of the mean equilibration time and the mean circuit complexity (for all operators, and averaged over initial conditions) yields a coefficient of determination of ${R^2 = 0.934}$, and an equivalent nonlinear model fit between natural logarithms of the mean equilibration time and mean classical entropy (i.e. the Shannon entropy of the equilibrium state vector, which can be computed by evaluating the moduli of the elements of the state vector, and then applying the Wolfram Language \textit{Entropy} function to the resultant list) yields a coefficient of determination of ${R^2 = 0.815}$.  Thus, we can conclude that there is indeed a strong exponential relationship between the quantum circuit complexity and the classical entropy, and a strong negative exponential relationship between both and the characteristic equilibration time (as was to be expected, both from the data visualized above, and from the conjectured analogy with classical statistical mechanics).

\subsection{Connections to Ergodic Theory and $k$-local Systems}

The observed equilibration phenomenon in QCAs can be explained as a direct consequence of von Neumann's mean ergodic theorem over arbitrary Hilbert spaces\cite{reed}\cite{vonneumann}\cite{vonneumann2}:

\begin{definition}
An \textit{orthogonal projection}, denoted $P$, in a Hilbert space ${\mathcal{H}}$ is any projection, that is, a linear operator satisfying:

\begin{equation}\nonumber
P^2 = P,
\end{equation}
for which the range and the null space are mutually orthogonal subspaces, i.e:

\begin{equation}\nonumber
\forall x, y \in \mathcal{H}, \qquad \bra{x}\ket{P y} = \bra{P x}\ket{P y} = \bra{P x}\ket{y}.
\end{equation}
\end{definition}

\begin{theorem}
If $U$ is a unitary operator on a Hilbert space ${\mathcal{H}}$, and $P$ is an orthogonal projection onto ${\ker{I - U}}$, then:

\begin{equation}\nonumber
\forall x \in \mathcal{H}, \qquad \lim_{N \to \infty} \frac{1}{N} \sum_{n = 0}^{N - 1} U^n x = P x.
\end{equation}
\end{theorem}

Thus, the mean ergodic theorem can be interpreted as stating that the sequences of averages of any unitary operator $U$:

\begin{equation}\nonumber
\frac{1}{N} \sum_{n = 0}^{N - 1} U^n,
\end{equation}
will always converge to the projection $P$, with respect to the strong operator topology. Consequently, the convergence of an arbitrary QCA state to an equilibrium state in the limit of a large number of applications of a global unitary evolution operator corresponds directly to this convergence of averages to the projection operator, where here the projection corresponds to the linear operator that maps any arbitrary state to the equilibrium state.

This allows us to make an explicit connection between the equilibration phenomenon investigated within this paper, and the ergodicity of $k$-local quantum systems, as analyzed in the context of black hole thermodynamics. As noted by Brown and Susskind, one might expect the phase space motion induced by a time-independent $k$-local Hamiltonian to be ergodic on ${SU(2^k)}$, but this conjecture can immediately be shown to be false by expressing the time-evolution operator in the energy basis:

\begin{equation}\nonumber
U = e^{- i H t} = \sum_n e^{- i E_n t} \ket{n} \bra{n}.
\end{equation}
For any given Hamiltonian, $U$ must therefore move on a phase space torus in ${2^k}$ dimensions (since there are ${2^k}$ energy eigenvalues), but the dimensionality of ${SU(2^k)}$ is ${4^k}$. However, motion on the ${2^k}$-torus is generally ergodic, since ergodicity is equivalent to the statement that the energy eigenvalues of the Hamiltonian are incommensurate, which will be true for almost any choice of weights.

\section{Concluding Remarks}

This paper has succeeded in its stated aim of introducing a novel mathematical formalism for QCAs, and using it to analyze and validate a variety of conjectures relating quantum computational complexity, classical thermodynamics, and black hole physics. Nevertheless, there clearly exist many possible directions for future research in the same vein.

One obvious direction is to improve the methods of complexity analysis. In particular, there is much scope for improvement of the coarse-grained approximations that we have used for analyzing the circuit complexity of the time reversal operator; indeed, the newly-released \textit{UniversalQCompiler} package\cite{iten} allows one to decompose arbitrary quantum operators directly into compositions of single-qubit rotations and CNOT gates, and could potentially be used to obtain dramatically better results than those presented here (though it would also require modifying our requirements for the time-reversal operator, by effectively enforcing unitarity rather than Hermiticity).

\begin{figure}[ht]
\includegraphics[width=0.95\textwidth]{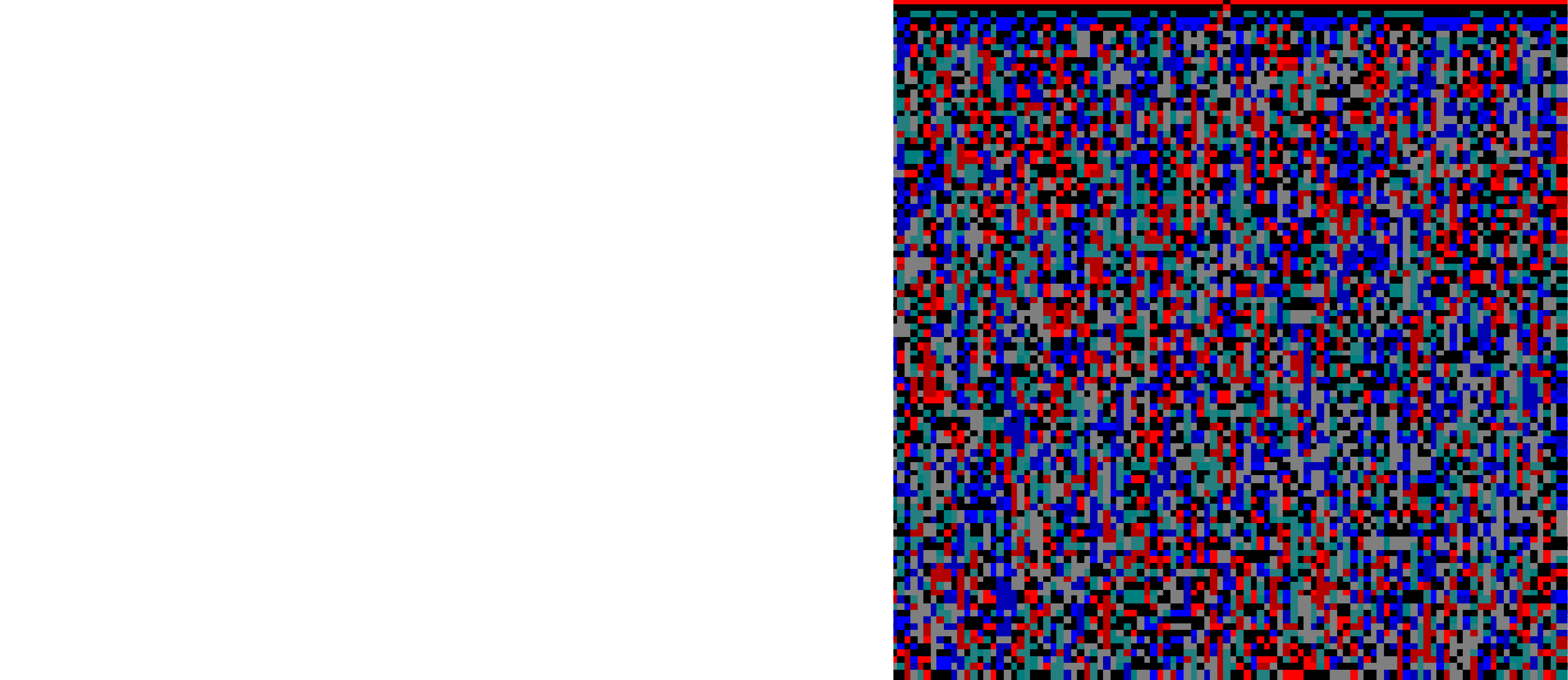}
\caption{An example run of a QCA based on the quantum teleportation protocol. The operator that gets applied to each qubit is either a Hadamard gate, a PauliX gate, a PauliY gate, or a PauliZ gate, depending upon whether the combined state of the teleported qubit and its entangled partner is ${\ket{00}}$, ${\ket{01}}$, ${\ket{10}}$ or ${\ket{11}}$, respectively.}
\label{figure11}
\end{figure}

A second obvious extension of this general research direction would be to repeat the same basic analyses (of both complexity-theoretic and thermodynamic QCA behavior) for a much larger class of QCA formalisms. For instance, one intermediate formulation that we investigated during the course of conducting this research was a QCA model based upon a quantum teleportation protocol, in which each qubit is updated by teleporting the neighboring qubit, and using the combined state of the teleported qubit and its entangled partner to determine which operator to apply to the original qubit. An example run of this QCA model is shown in Figure \ref{figure11}.

Ultimately, the qualitative behavior of these teleportation-based QCAs was deemed to be too similar to that of classical probabilistic CAs to be worthy of publication, though there doubtless exist many other QCA models for which the same conjectures explored within this paper (connecting classical entropy and quantum circuit complexity) can be analyzed and potentially validated.

Finally, there is a more conceptual direction to this project that may well be worthy of further exploration. The ``ER=EPR'' (Einstein-Rosen = Einstein-Podolsky-Rosen) conjecture in quantum gravity hypothesizes that all entangled particles may be connected by Einstein-Rosen bridges (wormhole solutions to the Einstein field equations)\cite{maldacena}, which implies more generally that spacetime may be ``made of entanglement'' in some fairly precise sense\cite{cowen}. More specifically, in the context of ER=EPR, it may become possible to link large-scale geometrical features of bulk spacetime to microscopic complexity-theoretic properties of the individual constituent quantum entanglements.

For instance, it is well-known that the neck of a classical Einstein-Rosen bridge grows with time (indeed, this may be a necessary condition for the consistency of general relativity, since it makes the wormhole effectively non-traversable), and it is also conjectured that the quantum computational complexity of a pair of entangled black holes will grow with time, by analogy with the second law of thermodynamics. In more precise terms, for a Schwarzschild black hole connecting two spacelike surfaces in $D$-dimensional anti-de Sitter space, with a metric of the form:

\begin{equation}\nonumber
ds^2 = -f(r) d \tau^2 + f(r)^{-1} dr^2 + r^2 d \Omega^{2}_{D - 2},
\end{equation}
where:

\begin{equation}\nonumber
f(r) = r^2 + 1 - \frac{\mu}{r^{D - 3}},
\end{equation}
and:

\begin{equation}\nonumber
\mu = 16 \pi G_N \frac{M}{(D - 2) \omega_{D - 2}},
\end{equation}
for ${D > 2}$, where ${\omega_{D - 2}}$ is the volume of a ${(D - 2)}$-sphere, the spatial volume of the Einstein-Rosen bridge is known to scale according to:

\begin{equation}\nonumber
\frac{d V}{d \tau} = \omega_{D - 2} r^{D - 2} \sqrt{\lvert f(r) \rvert}.
\end{equation}
Qualitatively, it seems reasonable that an Einstein-Rosen bridge will behave in a similar fashion to a pair of entangled black holes, and the more quantitative conjecture of Stanford and Susskind is that the quantum computational complexity of an entangled black hole state, denoted ${\mathcal{C}}$, will scale proportionally with the spatial volume of the associated Einstein-Rosen bridge:

\begin{equation}\nonumber
\mathcal{C} = \frac{V}{G_N l_{AdS}},
\end{equation}
where ${l_{AdS}}$ denotes the anti-de Sitter radius. We have already demonstrated that the QCA-based methods developed within this paper allow one to analyze quantitatively various complexity-theoretic conjectures regarding $k$-local systems (of exactly the kind used to model black holes), at least in toy cases. Therefore, there may well be the possibility for fruitful application of these techniques to special cases of the ER=EPR conjecture.

We fully intend to pursue these avenues of investigation, as well as many others, in future work.

\section*{Acknowledgments}
Both authors would like to thank Bentley University and the organizers of the 2019 Wolfram Summer School for their hospitality, and for facilitating the early stages of this research. They would also like to thank Stephen Wolfram for some stimulating conversations, and for initially suggesting the connection with ergodicity. RS thanks her parents. JG acknowledges financial support from the EPSRC under grant EP/L015552/1.

\end{document}